\documentstyle[twocolumn,aps,epsf]{revtex}

\begin{document}

\twocolumn[
\hsize\textwidth\columnwidth\hsize\csname @twocolumnfalse\endcsname

\title{Stable Elliptical Vortices in a Cylindrical Geometry}
\author{Peilong Chen}
\address{Department of Physics and Center for Complex Systems, 
         National Central University,
         Chungli 320, Taiwan}

\date{\today}

\maketitle
\begin{abstract}
We show that, in a two-dimensional (2d) ideal fluid (also applies to a
column of quasi-2d non-neutral plasma in an axial magnetic field),
large elliptical vortices in a finite disk are stable.  The stability
is established by comparison between energy of elliptical and
symmetrical states to satisfy a sufficient condition, without
dynamical eigen-analysis. Analytical small ellipticity expansion of
energy and exact numerical values for finite ellipticity are both
obtained.  The expansion indicates stable linear $l=2$ diocotron modes
for large vortices (or plasma columns). Numerical simulations of the
2d Euler equation are also performed. They not only confirm the
sufficient condition, but also show that the stability persists to
smaller vortex sizes.  The reason why decaying $l=2$ modes were
obtained by Briggs, Daugherty, and Levy {[Phys. Fluids {\bf 13}, 421
(1970)]} using eigen-analysis is also discussed.
\end{abstract}


\narrowtext
]

The two-dimensional (2d) incompressible Euler equation
\begin{equation}
  {\partial\omega\over\partial t} + ({\bf u}\cdot\nabla)\omega = 0,
  \label{eq:euler}
\end{equation}
not only describes an incompressible 2d ideal fluid, but also governs
the behavior of a long non-neutral plasma column confined by a uniform
axial magnetic field \cite{re:levy65}. Here ${\bf u}(x,y)$ is the 2d
velocity field and $\omega({\bf r})$ is the vorticity field,
$\omega\equiv(\nabla\times{\bf u})\cdot\hat{\bf z}$.  The
incompressibility condition, $\nabla\cdot{\bf u}=0$, can be
automatically satisfied by defining the stream function $\phi$ as
${\bf u} \equiv (\partial\phi/ \partial y,-\partial\phi/\partial x)$.
The stream function and vorticity are related by the Poisson equation
$\nabla^2\phi =-\omega$. In a pure electron plasma, $\omega$
corresponds to the electron density and $\phi$ to the electrical
potential.

Stability problems of coherent vortex states in this system are
long being interesting and important questions.  In a free space,
there exist exact nonlinear elliptical (Kirchoff) patch solutions
\cite{re:lamb32}.  In a cylindrical geometry Briggs, Daugherty, and
Levy \cite{re:briggs70} showed that, using dynamical eigen-analysis,
resonance between fluid elements and wave modes will
lead to damping of $l\ge2$ diocotron modes.  Here $l$ denotes the
mode number as the perturbation to a symmetric stream
function is written as $\phi_l(r)\exp[i(\Omega t-l\theta)]$.  By
solving the initial value problem of linearized equations and
properly treating analytical continuation in complex $\Omega$
plane, they obtained formulation for complex eigenvalue $\Omega$.
In particular, for a vorticity distribution very close to a step
function but negative radial derivative at all places, $\Omega$ with
a positive imaginary part is calculated for $l\ge2$, leading to
decaying normal modes.  

Experimental observations of decaying $l=2$ modes have been performed
by Pillai and Gould \cite{re:pillai94} in a pure electron plasma.
Exponential decay rates were obtained, as well as the observation of
fluid trapping in the diocotron mode at large amplitudes. In another
experiment with a pure electron plasma, \cite{re:mitchell94} beat-wave
resonance damping (transitions from high $l$ modes to low $l$ modes)
was observed to be the dominant vortex symmetrization mechanism.

A stability argument based on global constraints has also been
applied to the 2d vortex system \cite{re:holm85}. The logic of this
analysis is to show that a functional $W[\omega]$ which is conserved
by the 2d Euler equation is a maximum at a particular $\omega({\bf
r})$ against all other states that are accessible under incompressible
flows. At this maximum, no further changes in $\omega({\bf r})$ are
possible and the state is then stable.  For example, Davidson and Lund
\cite{re:davidson91} showed that a state in a cylindrical geometry
following a relation $\omega({\bf r}) =\omega(\phi({\bf r}))$ and
$\partial \omega(\phi)/ \partial\phi\ge 0$ is nonlinearly stable
\cite{re:rot}. In another example, O'Neil and Smith \cite{re:oneil92}
demonstrated that an off-center coherent vortex (linearly an $l=1$
perturbation) in a disk is also stable.  However, no results on the
stability of an $l=2$ mode using this method have been given in the
literatures.

Thermal equilibrium has been studied in 2d ideal fluids
\cite{re:mean_field}.  Since the coarse-grained entropy will not
decrease due to the dynamical vorticity mixing, it is proposed that
the system will reach a maximum coarse-grained entropy state at long
time.  Mean field equations governing these states have been derived
\cite{re:mean_field}, and solutions in some situations were obtained
\cite{re:mf_solutions}.  Once a mean-field equilibrium state is
obtained, its stability can be assured by showing a positive second
derivative of entropy against all possible perturbations. This of
course is using a similar principle as the method mentioned above.

In this paper we establish the stability of a large elliptical vortex
(comparing to the system size) against relaxation to a symmetrical
state using neither of the above two methods with eigen-analysis and
global maximum.  We will first deduce a stable sufficient condition
and then show that it is satisfied by elliptical vortices larger than
a critical radius.  The method is to compare energy of proper states,
not by evaluating second derivatives, actually not even finding any
equilibrium states. We further perform numerical simulations of the 2d
Euler equation to test our predictions.  Simulations not only
confirm the sufficient condition, but also show that elliptical
vortices are stable to lower radii.

Basic argument of the sufficient condition goes as: Consider initially
a uniform-vorticity elliptical vortex sitting at the center of a unit
disk, with unit vorticity level without losing generosity.  Now
consider its possible dynamics toward an axis-symmetrical vortex. This
will be a state with a linear $l=2$ diocotron mode if infinitesimal
ellipticity.

The Euler equation conserves the total vorticity $Q$, angular momentum
$M$, and energy $E$ of the initial ellipse, which are given by
\begin{eqnarray*}
  Q &=& \int\omega({\bf r})d{\bf r} \qquad
  M = \int r^2\omega({\bf r})d{\bf r} \\
  E_e &=& {\textstyle{1\over2}}\int\phi({\bf r})\omega({\bf r})d{\bf r}
\end{eqnarray*}
Furthermore, dynamical vorticity mixing ensures that the vorticity
level of the resulting symmetrical vortex will never exceed one (the
original uniform value). Under this restriction and given $Q$ and $M$
from the initial ellipse, there must be a maximum energy state with
its energy denoted as $E_s$ among all possible symmetrical
distributions.  With conservation of energy, this condition then
immediately follows:
\begin{quote}
   $E_e < E_s$ is necessary for the ellipse to ever evolve to a
   symmetrical vortex;

   $E_e > E_s$ is the {\em sufficient condition} for the ellipse {\em
   not} evolving to a symmetrical state.
\end{quote}
Applied to infinitesimal ellipticity, the $l=2$ diocotron mode
will not decay when $E_e > E_s$.

It should be noted here that this condition only try to exclude
symmetrical states from possible evolutions, a limitation purely
physically motivated.  For example, it seems unlikely that an ellipse
at the disk center will break the symmetry and relax to an off-center
vortex, although we believe that the energy of off-center vortices
could be larger than $E_e$. The conjecture (not decaying to off-center
states) is confirmed by numerical simulations which will be discussed
later.

To test the above condition, our first task is to calculate the energy
$E_e$ of a uniform elliptical vortex, which we define as a vorticity
distribution $\omega_e({\bf r})$ in the polar coordinate $(r,\theta)$
\begin{equation}
  \omega_e(r,\theta;r_0,\epsilon) = 1-s(r-r_0(1+\epsilon\cos2\theta)),
  \label{eq:omega_e}
\end{equation}
with $s(x)$ the usual step function.  The parameter $r_0$ defines a
base vortex size and $\epsilon$ its ellipticity. To compute the energy
of this vortex in a unit disk, first we consider the Green function in
a disk for the Poisson equation,
$
  \nabla^2 G({\bf r};{\bf r}') = -\delta({\bf r}-{\bf r}'),
$
with zero boundary condition at $r=1$.
Using an opposite-charged image charge sitting at ${\bf r}''\equiv
(1/r',\theta')$, the Green function can be written as
$
  G({\bf r};{\bf r}') = -{1\over 2\pi}\left(
    \ln|{\bf r}-{\bf r}'|-\ln|{\bf r}-{\bf r}''|-\ln r'\right).
$
The energy of the uniform elliptical vortex is then
\begin{eqnarray}
  E_e(r_0,\epsilon) 
    &=& {\textstyle{1\over2}}\int\phi({\bf r})\omega({\bf r})d{\bf r} 
               \nonumber\\
    &=& {\textstyle{1\over2}}
      \int_0^{2\pi}d\theta\int_0^{r_0(1+\epsilon\cos2\theta)}rdr \nonumber\\
    & & \int_0^{2\pi}d\theta'\int_0^{r_0(1+\epsilon\cos2\theta')}r'dr'
      G({\bf r};{\bf r}') \nonumber\\
    &=& E_0
      + \int_0^{2\pi}d\theta\int_{r_0}^{r_0(1+\epsilon\cos2\theta)}
       rdr\phi_0(r;r_0) \nonumber\\
    &+& {\textstyle{1\over2}}
       \int_0^{2\pi}d\theta\int_{r_0}^{r_0(1+\epsilon\cos2\theta)}rdr
              \label{eq:threeterms} \\
    & & \int_0^{2\pi}d\theta'\int_{r_0}^{r_0(1+\epsilon\cos2\theta')}r'dr'
      G({\bf r};{\bf r}'). \nonumber
\end{eqnarray}
We separate $E_e$ into three terms in the last equation.  Here
$\phi_0(r;r_0)$ is the stream function of a uniform circular vortex
with radius $r_0$, ${1\over r}{d\over dr}(r{d\phi_0\over
dr})=-\omega_0$, $\omega_0(r;r_0) = 1-s(r-r_0)$ and $E_0$ its
corresponding energy,
$$
  E_0= {\textstyle{1\over2}}
      \int\phi_0(r;r_0)\omega_0(r;r_0)2\pi rdr
     = \pi r_0^4\left(-{\textstyle{1\over4}}\ln r_0 
           + {\textstyle{1\over 16}}\right).
$$
We know of no way to integrate Eq. (\ref{eq:threeterms}) analytically.
Nevertheless, we can study the linear stability of an $l=2$ diocotron
mode from small $\epsilon$ behavior of $E_e(r_0,\epsilon)$. Since
the vortex is defined by $r_0(1+\epsilon\cos2\theta)$, the lowest
order dependence on $\epsilon$ must be $\epsilon^2$.  Correct to the
order of $\epsilon^2$, the second term in Eq. (\ref{eq:threeterms}) is
quickly found to be,
\begin{eqnarray*}
  & & \int_0^{2\pi}
       {\textstyle{1\over2}}\left[r_0\phi_0'(r_0;r_0)+\phi_0(r_0;r_0)\right]
       r_0^2\epsilon^2\cos^22\theta d\theta \\
  &=& -{\textstyle{1\over4}}\pi r_0^4(1+\ln r_0)\epsilon^2.
\end{eqnarray*}
Here prime denotes the derivative respected to $r$.  Evaluation of
the third term in Eq. (\ref{eq:threeterms}) is more difficult.  Again
correct to the order of $\epsilon^2$, the integration becomes
$$
  {\textstyle{1\over2}}r_0^4\epsilon^2
    \int_0^{2\pi}d\theta\int_0^{2\pi}d\theta'
    \cos2\theta\cos2\theta'G(r_0,\theta;r_0,\theta').
$$
Using the Green function and changing to new variables
$u\equiv\theta+\theta', v\equiv\theta-\theta'$, after some algebra,
we reach
\begin{equation}
  {\textstyle{1\over8}}r_0^4\epsilon^2
  \left[\pi+\int_0^{2\pi}\ln(a-\cos v)\cos2vdv\right],
  \label{eq:3rdterm}
\end{equation}
with $a\equiv {1\over2}(r_0^2+1/r_0^2)\ge 1$. The integration
$I\equiv\int_0^{2\pi}\ln(1-\cos v)\cos2vdv = -\pi$ has also been used
in reaching Eq. (\ref{eq:3rdterm}).

The integration in Eq. (\ref{eq:3rdterm}) is computed by first
integrating its derivative respective to $a$, and then using $I$ to
determine the constant arising from integration of $a$.  Eventually
Eq. (\ref{eq:3rdterm}) is found to be ${1\over8}\pi
r_0^4(1-r_0^4)\epsilon^2$, and the energy of the elliptical vortex becomes
\begin{equation}
  E_e(r_0,\epsilon) = E_0 + {\textstyle{1\over4}}\pi r_0^4\left(
  {-\textstyle{1\over2}}r_0^4-\textstyle{1\over2}-\ln r_0\right)\epsilon^2 
  +{\cal O}(\epsilon^4).
  \label{eq:energy_e}
\end{equation}

The energy $E_e(r_0,\epsilon)$ is now to be compared with the energy
$E_s$ of the maximum-energy symmetrical state with the same values of
total vorticity $Q$ and angular momentum $M$.  Its vorticity must also
be equal or less than one.  To see what this state is, first it is
favorable to have all the vorticity stay together, i.e., a uniform
unit-valued circular vortex with radius $r_s=(Q/\pi)^{1/2}$, to gain
as much as energy.  However this circular vortex has a fixed angular
momentum ${1\over2}\pi r_s^4$, and the uniform ellipse always has a
larger value. To satisfy the requirement of both $Q$ and $M$, as well
as achieving a maximum energy, the state will have a vorticity
distribution $\omega_s(r)$ as,
\begin{equation}
  \omega_s(r) = \left\{ \begin{array}{ll}
                 1 & \quad\hbox{for}\quad 0<r<\alpha\quad 
                                 \& \quad \beta<r<1 \\
                 0 & \quad\hbox{for}\quad \alpha<r<\beta.
                      \end{array}
              \right.
  \label{eq:omega_s}
\end{equation}
Here $\alpha$ and $\beta$ depend on $Q$ and $M$, which are determined
by $r_0$ and $\epsilon$.  In this distribution, a certain amount of
vorticity is put as far away from center as possible, i.e., at the
disk boundary, to account for the excess angular momentum and maximum
amount of vorticity is left to concentrate at the center to acquire a
maximum energy \cite{re:symmetrical}. Here we see how the system size
comes into play in a delicate manner.  At small $\epsilon$, $\alpha
= r_0\left(1+{1\over4}{1-3r_0^2\over 1-r_0^2}\epsilon^2\right)$ and
$\beta=1-{1\over2}{r_0^4\over1-r_0^2}\epsilon^2$, and the energy $E_s$
is expanded as (It involves only straightforward algebra to solve
$\phi_s$ and then integrate $E_s$.)
$$
  E_s = E_0 - {1\over4}\pi r_0^4{1-3r_0^2\over1-r_0^2}\ln r_0
  \epsilon^2 + {\cal O}(\epsilon^4).
$$
Now we obtain the energy difference between $E_e$ and $E_s$ as
$$
  E_e - E_s =
  {\pi\over4}r_0^4\left(-{1\over2}r_0^4-{1\over2}-
  {2r_0^2\over1-r_0^2}\ln r_0\right)\epsilon^2
  +{\cal O}(\epsilon^4).
$$
Evaluation of $\epsilon^2$ term reveals that there is a critical
value of $r_0$, $r_c\approx0.586$, such that $E_e<E_s$ for $r_0<r_c$
and $E_e>E_s$ for $r_0>r_c$.

So applying the energy condition, this indicates: the $l=2$ mode
perturbation of a circular vortex in a finite disk will not decay if
the vortex is large enough (larger than $0.586$ times the disk
radius).  This result seems contradict that of Briggs, Daugherty, and
Levy \cite{re:briggs70} where decaying modes were calculated from
eigen-analysis for all $l\ge2$ modes of a circular vortex with a
smooth profile very close to $\omega_0(r;r_0)$ (a step at $r_0$) but
negative $\omega'(r)$ at all $r$. The resolution lies at that in the
calculation of Briggs, Daugherty, and Levy, the symmetrical vortex is
assumed as a monotonic decreasing function of $r$. This seems a
reasonable and harmless condition.  However, as Eq. (\ref{eq:omega_s})
shows, this condition is very restrictive and always violated by the
uniform ellipse and hence their results no longer apply.

To further determine the stability of an ellipse with finite
ellipticity, we need go beyond the expansion and calculate the energy
for arbitrary $\epsilon$. Here we resort to numerical calculations of
integration $E={1\over2}\int\phi\omega d{\bf r}$. Since the Green
function using image charges has logarithmic functions and is not easy
to handle numerically, we rewrite the Green function as a summation of
Fourier components in the azimuthal direction,
$$
  G({\bf r};{\bf r}') = \sum_{m=0}^{\infty}g_m(r;r')
  \cos(m(\theta-\theta')),
$$
with $g_m$ power $\pm m$ of both $r$ and $r'$.  The energy now
becomes a summation on $m$ of four-dimensional $(r,\theta,r',\theta')$
integrals. The integration on $r$ and $r'$ can be carried out
analytically and the energy simplifies to a summation of double
integrals on $\theta$ and $\theta'$.  The integrals are then
calculated numerically, and results are checked to conform to
Eq. (\ref{eq:energy_e}) at small $\epsilon$.

The exact value of $E_e(r_0,\epsilon)$ now enable us to establish the
stability of finite ellipticity. In Figure 1 of the $r_0$--$\epsilon$
plane, we plot a solid line indicating the position where $E_e=E_s$.
(Although complicated, again $E_s$ with arbitrary $\epsilon$ can be
written down analytically from straightforward algebra). To the right
of the line, $E_e > E_s$ and an elliptical vortex will never relax to
a symmetrical vortex.  The line is almost vertical and only curves a
little to the left as $\epsilon$ is increased.  It crosses
$\epsilon=0$ at $r_0\approx0.586$, the value we have obtained from the
small $\epsilon$ expansion. To the left, the present analysis only
says that the decay to a symmetric state is allowed, but its
occurrence is not implied.

\begin{figure}
  \epsfxsize=3.25in
  \epsffile{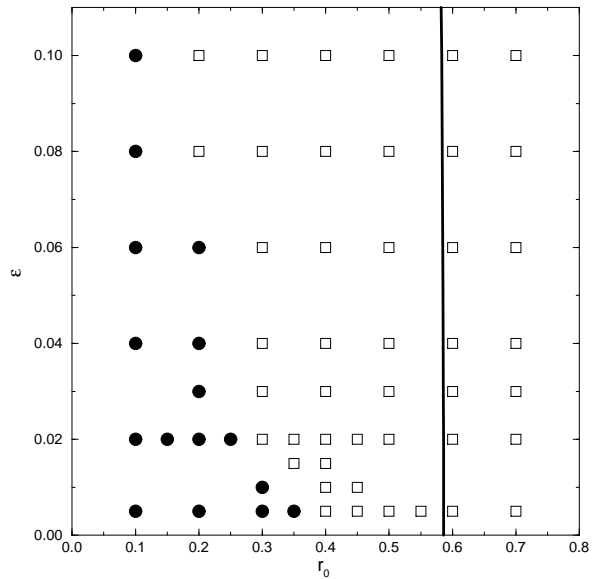}
  \caption{The vortex size and ellipticity space. The solid line marks
           the position where $E_e=E_s$. Squares represent relaxations
           to elliptical states in simulations, and circles to
           symmetrical states.}
\end{figure}

It should be emphasized here that we have proved that the ellipse
defined by Eq. (\ref{eq:omega_e}) will not decay to a symmetrical
state if $r_0>r_c$. It is very likely that dynamically it will undergo
adjustment and reach an elliptical-like steady state. We cannot say
about its exact distribution.  Current understanding is that it
probably should be a state described by $\omega({\bf r}) =
\omega(\phi({\bf r})+\Omega r^2)$, with $2\Omega$ giving the rigid
body rotation frequency around the disk center. With a particular
assumption on this functional dependence, an exact distribution can
then be computed. One example is the mean field equilibrium
\cite{re:mean_field}.  However, whether and when the system will reach
the prediction from this maximum-entropy principle (thermal
equilibrium) is still not very clear \cite{re:max}.

Although Eq. (\ref{eq:omega_e}) defines a uniform vortex, we do expect
that a smoothly distributed vortex should have similar stability
property if not deviated too much from Eq. (\ref{eq:omega_e}). The
exact values of stable radii will of course be different. This is
confirmed by the numerical simulations discussed next.

We also perform numerical simulations to test our predictions.
Simulations of the Euler equation in the polar coordinate have the
difficulty of singularity at the origin due to vanishing grid spacing.
To avoid this singularity, we use the functions,
\begin{eqnarray*}
  x &=& \mu\sqrt{1-\zeta^2/2} \\
  y &=& \zeta\sqrt{1-\mu^2/2},
\end{eqnarray*}
mapping a unit disk in the $x$--$y$ plane to a square in
$\mu$--$\zeta$ plane with $-1\le\mu\le1$ and $-1\le\zeta\le1$.  The
simulation is then done in $\mu$--$\zeta$ plane with Cartesian
coordinate.  The resolution is mostly $256\times256$, with a few
$512\times512$ runs to test convergence.  By avoiding the polar
coordinate and hence the singularity at the origin, we need a much
smaller numerical viscosity term, $\nu\nabla^2\omega$, to stabilize the
simulation and hence obtain more reliable long time results. It is
also noted that, since it is impossible to use a true step vorticity
distribution with finite grid points, the simulation results should
not be compared exactly with the predictions based on
Eq. (\ref{eq:omega_e}).

So for an initial ellipse with particular values of $r_0$ and
$\epsilon$, we run simulations to long time and determine their final
states. The results are plotted in Figure 1 as symbols, where squares
indicate relaxation to elliptical states and solid circles to
symmetrical states. Boundary region between squares and circles
represents the conditions where it is difficult to determine final
states from simulations.  In the figure we see the confirmation of
stable elliptical vortices with large vortex sizes.  All squares to
the right of the solid line shows that simulations are consistent with
the predictions.  The simulations also show that ellipses are actually
stable to a lower radius and the smallest stable size is decreasing
with increasing $\epsilon$. Finally no relaxations to off-center
vortices ever happen.

In conclusion we have showed that, from vorticity mixing in time
evolution and energy calculations, large elliptical vortices in a
finite disk will remain stable. At the infinitesimal ellipticity limit,
this indicates stable $l=2$ diocotron modes for large vortices.
Numerical simulations not only confirmed these results, but also shows
that elliptical states are actually stable to a smaller size.  The
contradiction to current general idea of decaying $l=2$ modes is also
indicated due to the incompleteness for considering only monotonic
decreasing vorticity by Briggs, Daugherty, and Levy.

The author thanks Dr. C. Y. Lu for fruitful discussions and C. R. Lo
of the help on numerical simulations. The support of National Science
Counsel, Taiwan, through the contract No. NSC 87-2112-M-008-034 is
also acknowledge.


\begin{references}

\bibitem{re:levy65} R. H. Levy, Phys. Fluids {\bf 8}, 1288 (1965).

\bibitem{re:lamb32} H. Lamb, {\em Hydrodynamics}, (Dover, New York, 1932)
                    6th ed., Secs. 158, 159.

\bibitem{re:briggs70} R. J. Briggs, J. D. Daugherty, and R. H. Levy,
                   Phys. Fluids {\bf 13}, 421 (1970).

\bibitem{re:pillai94} N. S. Pillai and R. W. Gould, Phys. Rev. Lett.
                      {\bf 73}, 2849 (1994).

\bibitem{re:mitchell94} T. B. Mitchell and C. F. Driscoll, Phys. Rev. Lett.
                      {\bf 73}, 2196 (1994).

\bibitem{re:holm85} Please see D. D. Holm, J. E. Marsden, T. Raitiu,
                    and A. Weinstein, Phys. Rep. {\bf 123}, 1 (1985)
                    for a review.  See also P. J. Morrison,
                    Rev. Mod. Phys. {\bf 70}, 467 (1998).

\bibitem{re:davidson91} R. C. Davidson and S. M. Lund,
                        Phys. Fluids B {\bf 3}, 2540 (1991).

\bibitem{re:rot} $\omega$ and $\phi$ here actually should be
                 defined in a rotating frame.

\bibitem{re:oneil92} T. M. O'Neil and R. A. Smith, 
                     Phys. Fluids B {\bf 4}, 2720 (1992).

\bibitem{re:mean_field} J. Miller, Phys. Rev. Lett. {\bf 65}, 2137
                        (1990); R. Robert and J. Sommeria, J. Fluid
                        Mech. {\bf 229} 291 (1991); J. Miller,
                        P. B. Weichman, and M. C. Cross, Phys. Rev. A
                        {\bf 45}, 2328 (1992).

\bibitem{re:mf_solutions} See, e.g., P. Chen and M. C. Cross, Phys. Rev.
                          E {\bf 54}, 6356 (1996).

\bibitem{re:symmetrical} Here we have not given a rigorous proof of
                         the maximum-energy symmetrical state. Please
                         see Ref. \cite{re:mf_solutions} for more
                         discussions about this state.

\bibitem{re:max} See, e.g., P. Chen and M. C. Cross, Phys. Rev. Lett.
                 {\bf 77}, 4174 (1996), and references therein.

\end{references}
\end{document}